# Estimating ground-level PM$_{2.5}$ by fusing satellite and station observations: A geo-intelligent deep learning approach

Tongwen Li [1], Huanfeng Shen [1,2,3], Qiangqiang Yuan [2,4], Xuechen Zhang [1], Liangpei Zhang [2,5]

[1] School of Resource and Environmental Sciences, Wuhan University, Wuhan, Hubei, 430079, China.

[2] The Collaborative Innovation Center for Geospatial Technology, Wuhan, Hubei, 430079, China.

[3] The Key Laboratory of Geographic Information System, Ministry of Education, Wuhan University, Wuhan, Hubei, 430079, China.

[4] School of Geodesy and Geomatics, Wuhan University, Wuhan, Hubei, 430079, China.

[5] The State Key Laboratory of Information Engineering in Surveying, Mapping and Remote Sensing, Wuhan University, Wuhan, Hubei, 430079, China.

Corresponding author: Huanfeng Shen (shenhf@whu.edu.cn)

Liangpei Zhang (zlp62@whu.edu.cn)

**Key Points:**

- A deep learning architecture is established to estimate ground-level PM$_{2.5}$ by fusing satellite and station observations.

- A geo-intelligent model is developed to incorporate geographical correlation into deep learning for performance improvement.

- This model shows a superior estimation accuracy (R=0.94, RMSE=13.68 $\mu g/m^3$) at national scale.

## Abstract

Fusing satellite observations and station measurements to estimate ground-level $PM_{2.5}$ is promising for monitoring $PM_{2.5}$ pollution. A geo-intelligent approach, which incorporates geographical correlation into an intelligent deep learning architecture, is developed to estimate $PM_{2.5}$. Specifically, it considers geographical distance and spatiotemporally correlated $PM_{2.5}$ in a deep belief network (denoted as Geoi-DBN). Geoi-DBN can capture the essential features associated with $PM_{2.5}$ from latent factors. It was trained and tested with data from China in 2015. The results show that Geoi-DBN performs significantly better than the traditional neural network. The cross-validation R increases from 0.63 to 0.94, and RMSE decreases from 29.56 to 13.68 $\mu g/m^3$. On the basis of the derived $PM_{2.5}$ distribution, it is predicted that over 80% of the Chinese population live in areas with an annual mean $PM_{2.5}$ of greater than 35 $\mu g/m^3$. This study provides a new perspective for air pollution monitoring in large geographic regions.

## 1. Introduction

$PM_{2.5}$, or particulate matter with an aerodynamic diameter of less than 2.5 $\mu m$, is associated with many adverse health effects, such as respiratory problems and cardiovascular disease [*Bartell et al.*, 2013]. Previous studies [*Y Chen et al.*, 2013] have shown that a 3-year reduction in average life expectancy and a 14% increase in overall mortality would result from a 100 $\mu g/m^3$ increase in the concentration of respirable particulate matter. As a result, $PM_{2.5}$ pollution has attracted widespread attention in recent years, and has become the focal point of international air pollution research.

With the launch of satellites and the continuous improvements in data retrieval technology, estimating ground-level $PM_{2.5}$ using satellite remote sensing has become a promising approach for the monitoring of $PM_{2.5}$ pollution. Three main kinds of methods have been applied to estimate $PM_{2.5}$ concentration using satellite-derived aerosol optical depth (AOD): chemical simulation models [*Liu et al.*, 2004; *van Donkelaar et al.*, 2010], statistical models [*Ma et al.*, 2016; *Song et al.*, 2014], and semi-empirical models [*Lin et al.*, 2015]. Among them, the statistical models are much easier to implement, and can obtain a competitive accuracy in $PM_{2.5}$ estimation [*Liu*, 2014]. As a result, many different statistical models have been developed to explore the quantitative relationship between satellite-derived AOD and ground-measured $PM_{2.5}$ (the AOD-$PM_{2.5}$ relationship). For example, the linear regression model establishes a simple linear relationship between AOD and $PM_{2.5}$. Considering more meteorological parameters, the multiple linear regression model was developed by *Gupta and Christopher* [2009a]. To account for the spatial heterogeneity of the AOD-$PM_{2.5}$ relationship, a geographically weighted regression model was introduced [*Hu et al.*, 2013]. Moreover, some more complex mixed-effect models [*Lee et al.*, 2011] and generalized additive mixed models [*Kloog et al.*, 2011] have also been developed to estimate ground-level $PM_{2.5}$. All these statistical models are used to represent the relationship between $PM_{2.5}$ and the latent factors.

However, the levels of $PM_{2.5}$ concentration are related to many factors, such as meteorological conditions (e.g., temperature, wind speed, relative humidity), land-use type, population, road networks, and so on. This situation has increased the difficulty of using the traditional statistical models to estimate $PM_{2.5}$. Unlike the traditional methods, intelligent algorithms have the capacity to work better with this problem. For example, *Gupta and*

*Christopher* [2009b] used a back-propagation neural network (BPNN) to estimate surface-level $PM_{2.5}$ in the southeastern United States; an artificial neural network algorithm was trained with Bayesian regularization to estimate $PM_{2.5}$ in eastern China [*Y Wu et al.*, 2012]; and a generalized regression neural network (GRNN) was reported to outperform the traditional models at national scale in China [*Li et al.*, 2017]. These neural network models show great advantages in estimating ground-level $PM_{2.5}$ concentrations.

Deep learning, which is considered to be the second generation of neural network, may be a potential way to address this situation [*Hinton and Salakhutdinov*, 2006]. However, to date, deep learning has seldom been applied in the estimation of ground-level $PM_{2.5}$, only a few attempts [*Ong et al.*, 2015] have been made to predict time-series $PM_{2.5}$ concentrations over monitoring stations. On the other hand, intelligent algorithms are usually used to describe numerical relationships, but they neglect the geographical correlation of environmental variables. Meanwhile, it has been reported that $PM_{2.5}$ concentrations show significant autocorrelation in time and space [*J Wu et al.*, 2015]. The nearby $PM_{2.5}$ from neighboring stations and the $PM_{2.5}$ observations from nearby days for the same station are informative for estimating $PM_{2.5}$. It is therefore important to incorporate this geographical correlation relationship into the intelligent algorithms.

Consequently, the objective of this study is to develop a geo-intelligent deep learning (Geoi-DL) model to estimate ground-level $PM_{2.5}$ concentrations. This model is established and evaluated based on satellite observations, meteorological parameters, and ground-level $PM_{2.5}$ measurements from China, which is suffering from serious $PM_{2.5}$ pollution [*Che et al.*, 2007; *Peng et al.*, 2016]. This study will provide a new perspective to investigate the spatiotemporal characteristics of air pollution in a large geographic region.

## 2. Materials and Methods

### 2.1. Study Region and Data

The study region is China (Figure S1). The study period is from January 1, 2015, to December 31, 2015, with a total length of 365 days.

The data used include four main parts. **1) Ground-level $PM_{2.5}$ data**. Hourly $PM_{2.5}$ data for 2015 were obtained from the China National Environmental Monitoring Center (CNEMC) website (http://www.cnemc.cn). The number of monitoring stations was ~1500 by the end of 2015. We averaged hourly $PM_{2.5}$ to daily mean $PM_{2.5}$ for the estimation of $PM_{2.5}$. **2) MODIS AOD.** Both Terra and Aqua MODIS Level 2 AOD products were downloaded from the Level 1 and Atmosphere Archive and Distribution System (LAADS, http://ladsweb.nascom.nasa.gov). We adopted Collection 6 (C6) 10-km AOD products, which are retrieved by combining dark target and deep blue algorithms [*Levy et al.*, 2013]. The average of the Terra and Aqua AOD products was employed to estimate daily average $PM_{2.5}$. **3) Meteorological parameters.** We extracted relative humidity (RH, %), air temperature at a 2 m height (TMP, K), wind speed at 10 m above ground (WS, m/s), surface pressure (PS, Pa), and planetary boundary layer height (PBL, m) from MERRA-2 meteorological reanalysis data, which were downloaded from the website (http://gmao.gsfc.nasa.gov/GMAO_products/). **4) MODIS normalized difference vegetation index (NDVI).** MODIS NDVI products (Level 3, MOD13) are available at a resolution of 1 km every 16 days, and were downloaded from the LAADS website. MODIS NDVI was incorporated into the AOD-$PM_{2.5}$ model to reflect the land-use type.

We created a 0.1-degree grid for the data integration and model establishment. All the data were reprocessed to be consistent temporally and spatially, to form a complete dataset. Ground-level PM$_{2.5}$ data observed from multiple stations in each grid were averaged. The satellite-derived AOD, NDVI, and meteorological reanalysis data were regridded to 0.1 degrees and reprojected to the same projection coordinate system. After the data preprocessing and integration, a total of 71084 records were collected for the model development.

## 2.2. The Deep Belief Network (DBN) Model for the Estimation of PM$_{2.5}$

The DBN model is one of the most typical deep learning models, and it was introduced in 2006 [*Hinton et al.*, 2006]. A DBN consists of multiple restricted Boltzmann machine (RBM) layers and a back-propagation (BP) layer, which can be used for classification or prediction problems. For example, the structure of a DBN with two RBM layers is shown in Figure 1(a).

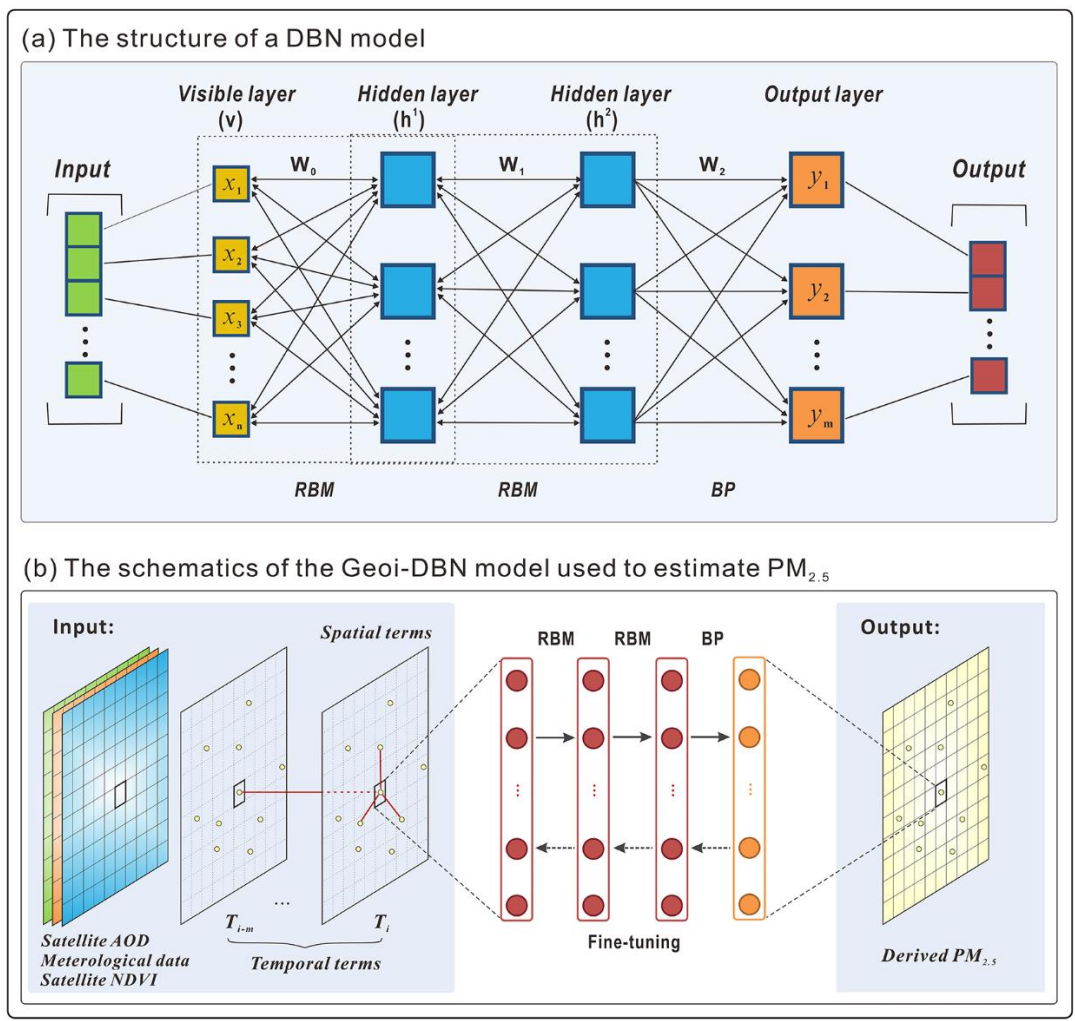

**Figure 1**. The structure of a DBN and the specific schematics (Geoi-DBN) used to estimate PM$_{2.5}$.

An RBM consists of a visible layer and a hidden layer, where the hidden layer of the prior RBM is the visible layer of the next RBM. Taking the first RBM as an example, from the visible layer ($\mathbf{v}$) to the hidden layer ($\mathbf{h}^1$),

$$h_i^1 = \begin{cases} 1, & f(\mathbf{W}_{0,i}\mathbf{x}+b_i) \geq \mu \\ 0, & f(\mathbf{W}_{0,i}\mathbf{x}+b_i) < \mu \end{cases} \qquad \mu \sim U(0,1) \tag{1}$$

where $i$ refers to the number of the *ith* neuron, and $b_i$ denotes the bias for neuron $i$. $f(\cdot)$ indicates the transfer function, $f(x) = \frac{1}{1+e^{-x}}$. It is the same for the calculation of the visible layer from the hidden layer. The contrastive divergence algorithm is usually used for training an RBM [*Hinton and Salakhutdinov*, 2006]. The weights are updated in the *nth* iteration as:

$$\mathbf{W}_0^{n+1} = \mathbf{W}_0^n + \varepsilon \cdot \left( (\mathbf{h}_1^1)^T \mathbf{x} - (\mathbf{h}_2^1)^T \mathbf{v}_1 \right) \tag{2}$$

where $\varepsilon$ is the learning rate, $\mathbf{v}_1$ denotes the reconstruction from hidden layer ($\mathbf{h}_1^1$), and $\mathbf{h}_1^1, \mathbf{h}_2^1$ are generated from $\mathbf{x}, \mathbf{v}_1$ using Equation (1), respectively. The RBMs are pre-trained one by one, without supervision, and the trained weights are used to initialize the multi-layer neural networks. The DBN model then works as a feed-forward neural network [*Yue et al.*, 2017], whereas the error reduction using the BP algorithm is referred to here as "fine-tuning".

Specifically, in our case, the schematics of the geo-intelligent DBN (Geoi-DBN) used to estimate ground-level $PM_{2.5}$ are presented in Figure 1(b). The input variables are the satellite-derived AOD, meteorological parameters, NDVI, and spatiotemporally informative terms. Because of the autocorrelation, the nearby $n$ grids of $PM_{2.5}$ measurements and the $PM_{2.5}$ observations from the $m$ prior days for the same grid are informative for estimating $PM_{2.5}$. The nearer observations are more informative than further ones [*Tobler*, 1970; *Yuan et al.*, 2012]. For a specific grid, the spatiotemporally informative terms are represented as:

$$S\text{-}PM_{2.5} = \frac{\sum_{i=1}^{n} ws_i PM_{2.5,i}}{\sum_{i=1}^{n} ws_i} \qquad ws_i = \frac{1}{ds_i^2} \tag{3}$$

$$T\text{-}PM_{2.5} = \frac{\sum_{j=1}^{m} wt_j PM_{2.5,j}}{\sum_{j=1}^{m} wt_j} \qquad wt_j = \frac{1}{dt_j^2} \tag{4}$$

$$DIS = \min\left(\frac{1}{ds_i}\right) \quad i = 1, 2, ..., n \tag{5}$$

where $ds, dt$ refer to the spatial and temporal distances, respectively. $m, n$ are 3 and 10, respectively. The geographical distance ($DIS$) is used to reflect the heterogeneity of uneven station distribution. Two hidden layers (two RBMs) are then used, and the number of neurons in each hidden layer is 15 (Text S1 in the supporting information). The RBM layers are stacked one by one to transfer the input signals to the higher layer. The output layer is a BP layer, which has only one node ($PM_{2.5}$ measurements).

The relationship $PM_{2.5} = f(AOD, RH, WS, TMP, PBL, PS, NDVI, S\text{-}PM_{2.5}, T\text{-}PM_{2.5}, DIS)$ is wished to learn from the data records. The process can be divided into three steps:

**1) Pre-training**. Using the collected data records, the RBMs are trained layer by layer, without supervision. This unsupervised training can extract the essential features associated with PM$_{2.5}$, and they are transferred from the prior RBM to the next RBM layer. Therefore, the higher layer can extract the deeper features related to PM$_{2.5}$.

**2) Fine-tuning**. Through the prior pre-training step, the initial weights of Geoi-DBN are generated and we can obtain the calculated PM$_{2.5}$. Compared with in-situ PM$_{2.5}$ measurements, an estimation error can be obtained, and it is sent back to the Geoi-DBN model to fine-tune the weight coefficients using the BP algorithm.

**3) Prediction**. This step evaluates the performance of the Geoi-DBN model established on the input data records, and predicts the PM$_{2.5}$ values for those locations with no ground stations. Thus, spatially continuous PM$_{2.5}$ data can be reconstructed.

Furthermore, to evaluate the model performance, a 10-fold cross-validation technique [*Rodriguez et al.*, 2010] was applied to test the model overfitting and predictive power. We adopted the statistical indicators (Text S2 in the supporting information) of the correlation coefficient (R), the root-mean-square error (RMSE, $\mu g/m^3$), the mean prediction error (MPE, $\mu g/m^3$), and the relative prediction error (RPE, defined as RMSE divided by the mean ground-level PM$_{2.5}$) to evaluate the model performance.

## 3. Results and Discussion

### 3.1. Evaluation of the Model Performance

**Table 1.** The performance of the models

| Model | Model fitting | | | | Cross-validation | | | |
|---|---|---|---|---|---|---|---|---|
| | R | RMSE | MPE | RPE (%) | R | RMSE | MPE | RPE (%) |
| Ori-BPNN | 0.66 | 28.58 | 19.97 | 52.57 | 0.63 | 29.56 | 20.74 | 54.35 |
| Ori-GRNN | 0.88 | 18.51 | 12.64 | 34.02 | 0.78 | 23.82 | 16.34 | 43.79 |
| Ori-DBN | 0.76 | 25.49 | 18.02 | 46.18 | 0.74 | 26.33 | 18.42 | 47.62 |
| Geoi-BPNN | 0.92 | 15.23 | 10.16 | 27.65 | 0.91 | 15.74 | 10.62 | 28.65 |
| Geoi-GRNN | 0.93 | 14.66 | 11.04 | 26.70 | 0.90 | 17.40 | 12.66 | 31.67 |
| Geoi-DBN | 0.94 | 13.44 | 8.91 | 24.50 | 0.94 | 13.68 | 9.03 | 24.90 |

To evaluate the deep learning model, we compared the DBN model with BPNN [*Gupta and Christopher*, 2009b] and GRNN [*Li et al.*, 2017]. The comparison of the model performance is presented in Table 1. When the geographical correlation is not incorporated into these models (original models), Ori-GRNN obtains the best performance (cross-validation R=0.78, RMSE=23.82 $\mu g/m^3$), while Ori-BPNN performs the worst. The results agree with our previous study [*Li et al.*, 2017]. The Ori-DBN model shows a worse performance than Ori-GRNN. The reason for this could be that the AOD-PM$_{2.5}$ relationship is not complicated enough for deep learning. From the original models to the geo-intelligent models, the performance is greatly improved. However, it is worth noting that among these models, the Geoi-DBN model performs the best, whereas the Geoi-GRNN model performs the worst. A possible reason for this is that the spatiotemporally informative terms greatly increase the complexity of the AOD-PM$_{2.5}$ relationship. Benefiting from more layers and layer-by-layer pre-training, the much more complicated relationship between PM$_{2.5}$ and the predictors is better learned in Geoi-DBN than in Geoi-GRNN. Therefore, the Geoi-DBN model achieves the best performance, with cross-validation R and RMSE values of 0.94 and 13.68 $\mu g/m^3$, respectively. These results demonstrate that the Geoi-DBN model, which considers the geographical correlation, is a promising approach for describing the AOD-PM$_{2.5}$ relationship.

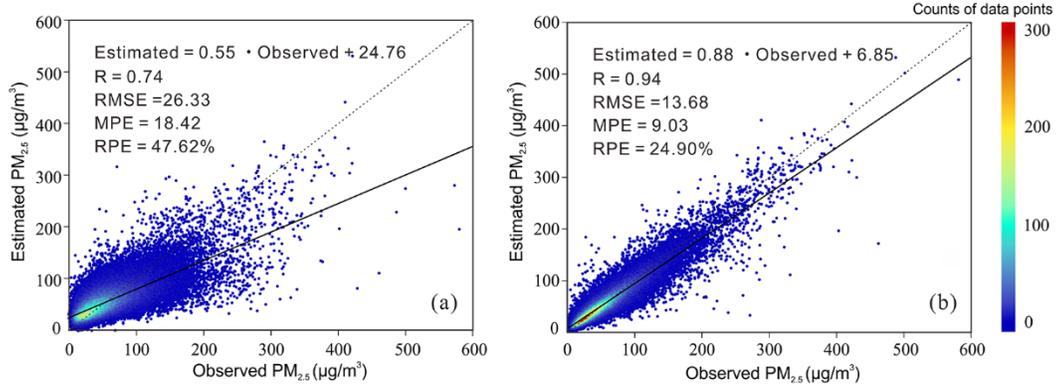

**Figure 2.** Scatter plots of the cross-validation results. (a) Cross-validation results of the Ori-DBN model. (b) Cross-validation results of the Geoi-DBN model. The dashed line is the $y = x$ line as reference.

Figure 2 shows the scatter plots for cross-validation of the Ori-DBN and Geoi-DBN models. For the Ori-DBN model, the cross-validation R and RMSE values are 0.74 and 26.33 $\mu g/m^3$, respectively. The R and RMSE values of the model fitting are 0.76 and 25.49 $\mu g/m^3$ (see Figure S2), respectively. When considering the geographical correlation, the spatiotemporal characteristics of atmospheric PM$_{2.5}$ are better described in the AOD-PM$_{2.5}$ modeling. The model performance is therefore significantly improved. On the other hand, the cross-validation slope of observed PM$_{2.5}$ versus prediction for the Geoi-DBN model is 0.88, with an intercept of 6.85 $\mu g/m^3$. These findings indicate that, despite the good fitting with high R values, the Geoi-DBN model tends to underestimate when the ground-level PM$_{2.5}$ is greater than ~60 $\mu g/m^3$. Therefore, the higher PM$_{2.5}$ concentrations may not be sufficiently explained. This issue is further discussed in Section 3.4. However, it should be noted that the cross-validation slope of the Geoi-DBN model (0.88) is much greater than that of the Ori-DBN model (0.55). This means that the Geoi-DBN model shows a much lower extent of underestimation than the Ori-DBN model.

### 3.2. Mapping of PM$_{2.5}$ Concentrations

In Figure 3, the annual mean distribution of PM$_{2.5}$ concentrations in China is mapped, based on the Geoi-DBN model and our previous mapping strategy [*Li et al.*, 2017]. Overall, the levels of PM$_{2.5}$ concentrations are higher in the northern regions than the southern regions. Meanwhile, a heavily polluted region is located in the North China Plain. As reported in a previous study [*Z Chen et al.*, 2008], the climate of this region is characterized by stagnant weather, with weak wind and a relatively low boundary layer height, which results in the atmospheric conditions for the accumulation, formation, and processing of aerosols. This is one of the main reasons for the serious PM$_{2.5}$ pollution in this area. Additionally, the PM$_{2.5}$ concentrations are generally higher in the inland regions (e.g., Hunan, Hubei, and Hunan provinces), and lower in the coastal regions (e.g., Guangdong and Fujian provinces). The regions with the least PM$_{2.5}$ pollution are in Hainan and Yunnan provinces, which benefit from the low levels of anthropogenic emissions and favorable meteorological conditions for atmospheric dispersion. Last but not least, a very high level of PM$_{2.5}$ pollution is found in the northwest, especially the Xinjiang Autonomous Region. A possible reason for this is that the dust particles in this desert region make a significant contribution to the accumulation of PM$_{2.5}$ [*Fang et al.*, 2016].

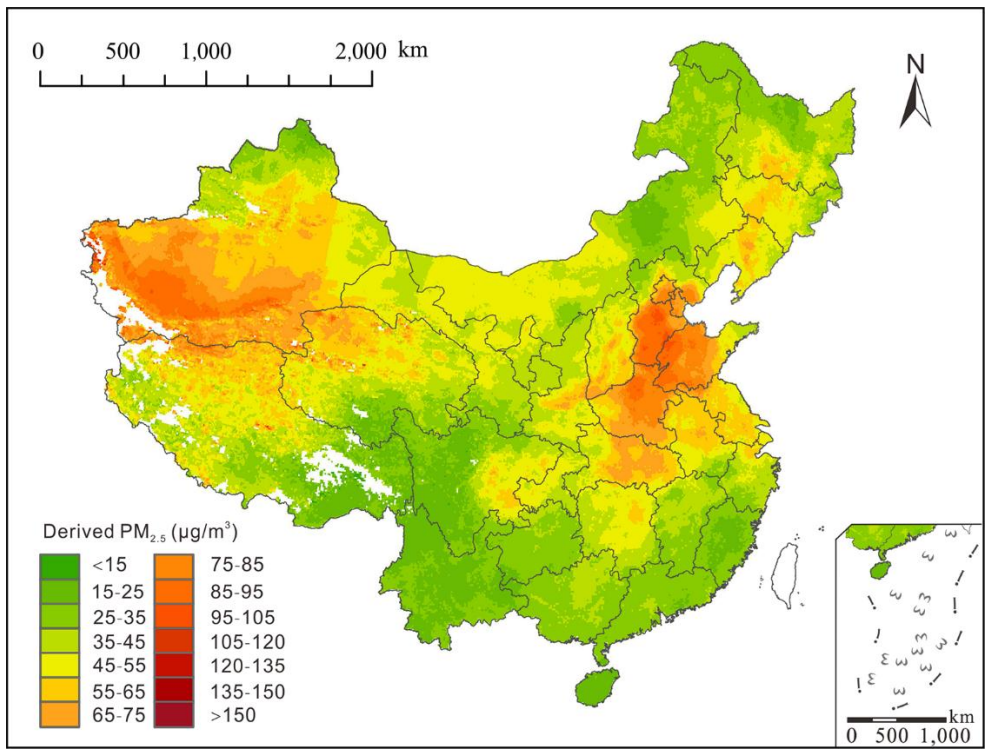

**Figure 3.** Annual mean distribution of PM$_{2.5}$ in China. The white regions indicate missing data.

The mean seasonal distributions of PM$_{2.5}$ in China are shown in Figure S3. High levels of PM$_{2.5}$ concentrations occur in winter, while the summer shows the lowest PM$_{2.5}$ levels. It is also worth noting that spatial variability is also found, especially in Northwest China. The PM$_{2.5}$ pollution in this region is very serious during spring, but is much decreased in autumn. The reason for this is that dust storms frequently occur in this region in spring in the desert, semi-desert, and grassland areas [*Zou and Zhai*, 2004].

### 3.3. Exposure Analysis over China

The population data were obtained from the Socioeconomic Data and Applications Center (SEDAC, http://sedac.ciesin.columbia.edu/data/collection/gpw-v4). The distribution of population in China in 2015 is presented in Figure S4. The World Health Organization (WHO) interim targets (IT)-1 and IT-3 for annual mean PM$_{2.5}$ concentration are 35 and 15 $\mu g/m^3$, respectively [*WHO*, 2006]. As shown in Figure 4, the population-weighted estimated annual mean PM$_{2.5}$ is 53.44 $\mu g/m^3$, which exceeds the WHO IT-1 standard. Almost all regions of

China (except for the Northwest and South) show population-weighted averages that are greater than the spatial averages. These findings indicate that more people are living in relatively more polluted regions. Figure 4 also shows that over 80% of the Chinese population live in areas that exceed the WHO IT-1 standard. Spatially, South China has the highest percentage of population living in areas meeting the WHO IT-1 standard, whereas Central and North China have the lowest. These findings demonstrate that China is still suffering from serious $PM_{2.5}$ pollution, and more attention needs to be paid to $PM_{2.5}$ pollution.

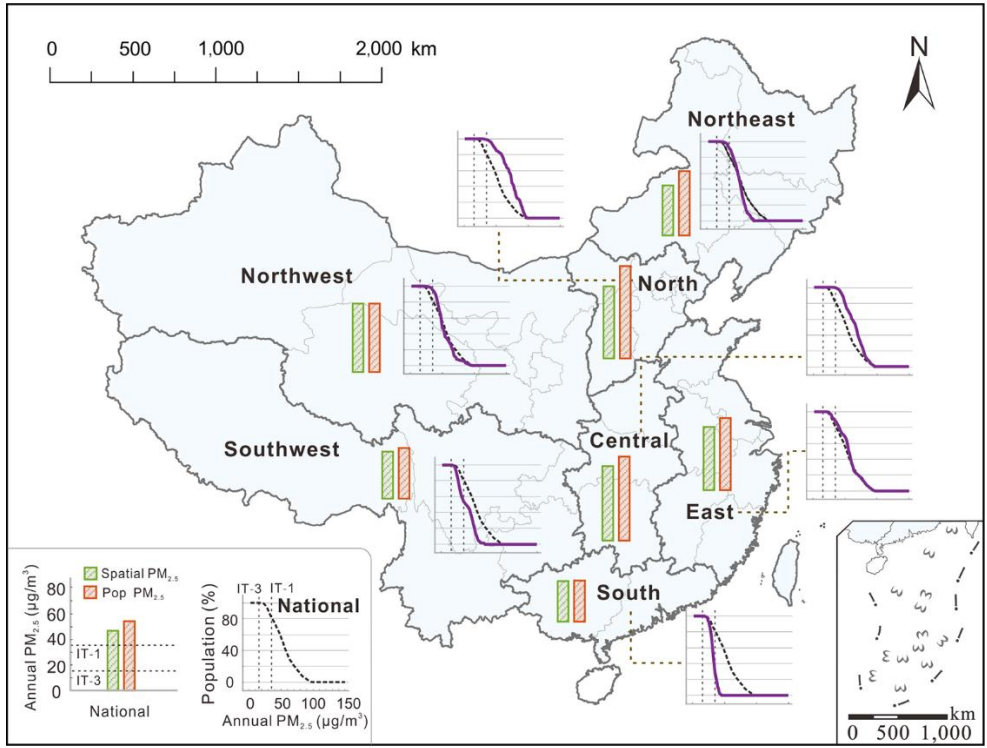

**Figure 4.** Exposure to $PM_{2.5}$ over China in 2015. Spatial $PM_{2.5}$: spatial mean $PM_{2.5}$. Pop $PM_{2.5}$: population-weighted mean $PM_{2.5}$. The curves represent the percentage of population exposed to $PM_{2.5}$, and the black dashed curve denotes national exposure to $PM_{2.5}$.

### 3.4. Discussion

$PM_{2.5}$ assessment by fusing satellite and station observations involves lots of different factors, which inherently results in big data. In this situation, the deep learning model may better estimate $PM_{2.5}$. Therefore, we made efforts to incorporate urban big data (i.e., road network and population data) which were used in a previous study [*Fang et al.*, 2016] into the deep learning model. The data were acquired from the National Geomatics Center of China (NGCC, http://ngcc.sbsm.gov.cn/) and SEDAC, respectively. The results show that these predictors have almost no positive (or even passive) effect on model performance (R=0.94, RMSE=13.67 $\mu g/m^3$, see Table S2). A possible reason could be that these predictors are not real-time, and cannot reflect the temporal variation of the AOD-$PM_{2.5}$ relationship. It is possible that the model performance would be greatly improved if real-time data (e.g., daily traffic flow) were obtained.

The Geoi-DBN cross-validation slope of observed $PM_{2.5}$ versus estimation is 0.88, indicating some evidence of bias. However, it should be noted that the national-scale estimates of $PM_{2.5}$ in China [*Fang et al.*, 2016; *Lin et al.*, 2015; *Ma et al.*, 2016; *You et al.*, 2016] are mostly underestimated (slopes of 0.79~0.83) when ground-level PM$_{2.5}$ is greater than 60 $\mu g/m^3$. This underestimation may be down to several reasons, including the possibility of mixed layers of aerosols in the atmosphere, and the hygroscopicity of urban aerosols [*Gupta and Christopher*, 2009b]. Therefore, this underestimation is probably a systematic error related to the complicated aerosols in China.

In this study, a deep learning architecture has been established to estimate ground-level PM$_{2.5}$, achieving a satisfactory performance. However, it should be noted that we applied only one type of deep learning model (i.e., DBN) to model the AOD-PM$_{2.5}$ relationship. Would any other deep learning model work better with this problem? Deep learning has more hidden layers to better represent complex non-linear relationships, so whether or not we can estimate PM$_{2.5}$ using original satellite reflectance rather than satellite-derived AOD to avoid intermediate error deserves further study. Deep learning is a promising approach for AOD-PM$_{2.5}$ modeling, but there is still room for improvement.

## 4. Conclusions

Despite the potential application of satellite-based AOD for air quality studies [*Wang and Christopher*, 2003], the estimation of PM$_{2.5}$ concentrations involves a large number of factors. We therefore developed a geo-intelligent deep learning model to better represent the AOD-PM$_{2.5}$ relationship. This study introduced the layer-by-layer pre-training technique to the satellite remote sensing assessment of PM$_{2.5}$. In addition, the geographical correlation was adopted to significantly improve the estimation accuracy. The deep learning-based AOD-PM$_{2.5}$ modeling of China accurately estimated PM$_{2.5}$ concentrations, with cross-validation R and RMSE values of 0.94 and 13.68 $\mu g/m^3$, respectively. It is predicted that over 80% of the Chinese population live in areas with an annual mean PM$_{2.5}$ greater than the WHO IT-1 standard (35 $\mu g/m^3$) in 2015. Overall, we can say that the proposed approach is promising for air pollution monitoring in large geographical regions.

**Acknowledgments**

This work was funded by the National Key R&D Program of China (2016YFC0200900) and the National Natural Science Foundation of China (41422108). We are grateful to the China National Environmental Monitoring Center (CNEMC), the Goddard Space Flight Center Distributed Active Archive Center (GSFC DAAC), the US National Aeronautics and Space Administration (NASA) Data Center, the National Geomatics Center of China (NGCC), and the Socioeconomic Data and Applications Center (SEDAC) for providing the foundational data for free.

Supporting Information for

# Estimating ground-level PM$_{2.5}$ by fusing satellite and station observations: A geo-intelligent deep learning approach

Tongwen Li [1], Huanfeng Shen [1,2,3], Qiangqiang Yuan [2,4], Xuechen Zhang [1], Liangpei Zhang [2,5]

[1] School of Resource and Environmental Sciences, Wuhan University, Wuhan, Hubei, 430079, China.

[2] The Collaborative Innovation Center for Geospatial Technology, Wuhan, Hubei, 430079, China.

[3] The Key Laboratory of Geographic Information System, Ministry of Education, Wuhan University, Wuhan, Hubei, 430079, China.

[4] School of Geodesy and Geomatics, Wuhan University, Wuhan, Hubei, 430079, China.

[5] The State Key Laboratory of Information Engineering in Surveying, Mapping and Remote Sensing, Wuhan University, Wuhan, Hubei, 430079, China.

**Contents of this file**

    Text S1 to S2
    Figures S1 to S4
    Tables S1 to S2

**Introduction**

The supporting information provides details of the selection of numbers of RBM layers and neurons in the DBN model, and calculation of the statistical indictors. In addition, the fitting results of the Ori- and Geoi-DBN models, the distribution of the Chinese population, and the seasonal mean distributions of PM$_{2.5}$ in China are presented. We also provide details of the influence of predictors (road and population) on the Geoi-DBN model.

**Text S1. Selection of the Numbers of RBM Layers and Neurons in the DBN Model**

The number of RBM layers and the numbers of neurons in the RBM layers are the critical parameters in DBN modeling. Generally speaking, more RBM layers in the DBN model are suggested to model abstract concepts and complex relationships; however, this is more likely to result in overfitting of the input data [*Le Roux and Bengio*, 2008]. It has been shown that the neural network performs better with the number of neurons in the range of $\left(2\sqrt{n}+\mu, 2n+1\right)$ [*Fletcher and Goss*, 1993], where $n$ is the number of input variables, and $\mu$ is the number of output variables. As shown in Table S1, there are no significant differences between the various parameters in our situation. One RBM layer appears to result in a slight advantage; however, it should be noted that this study involves complex atmospheric processes. Therefore, we adopted two RBM layers with 15 neurons in each RBM layer, to make a compromise between the model fit, the predictive power, and the computation time.

**Text S2. Calculation of Statistical Indicators**

We adopted the statistical indicators of the correlation coefficient (R), root-mean-square error (RMSE, $\mu g/m^3$), mean prediction error (MPE, $\mu g/m^3$), and relative prediction error (RPE, %) to evaluate the model performance. These indicators are calculated as follows:

$$R = \frac{\sum_{i=1}^{n}\left(PM_{o,i} - \overline{PM_o}\right)\left(PM_{e,i} - \overline{PM_e}\right)}{\sqrt{\sum_{i=1}^{n}\left(PM_{o,i} - \overline{PM_o}\right)^2}\sqrt{\sum_{i=1}^{n}\left(PM_{e,i} - \overline{PM_e}\right)^2}} \tag{S1}$$

$$RMSE = \sqrt{\frac{\sum_{i=1}^{n}\left(PM_{o,i} - PM_{e,i}\right)^2}{n}} \tag{S2}$$

$$MPE = \frac{\sum_{i=1}^{n}\left|PM_{o,i} - PM_{e,i}\right|}{n} \tag{S3}$$

$$RPE = \frac{RMSE}{\overline{PM_o}} \tag{S4}$$

where $n$ is the total number of data records, and $PM_o, PM_e$ are the ground-observed PM$_{2.5}$ and model-estimated PM$_{2.5}$, respectively.

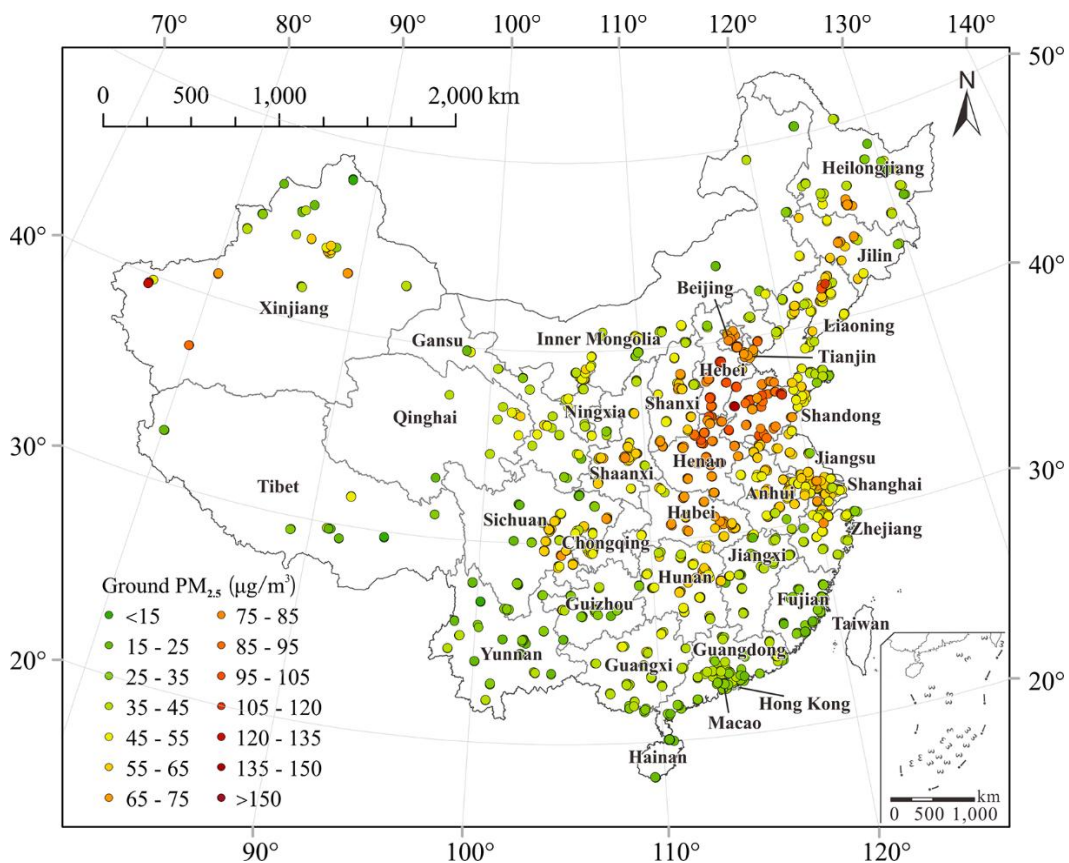

**Figure S1.** Study region and spatial distribution of the PM$_{2.5}$ monitoring stations. The annual mean PM$_{2.5}$ in 2015 was calculated from the measurements over these stations.

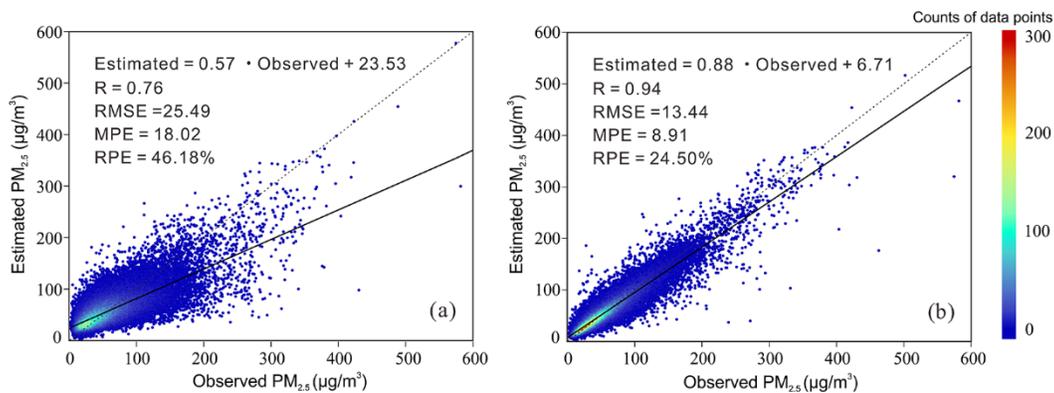

**Figure S2.** Scatter plots of the model fitting results. (a) Fitting results of the Ori-DBN model. (b) Fitting results of the Geoi-DBN model. The dashed line is the $y = x$ line as reference.

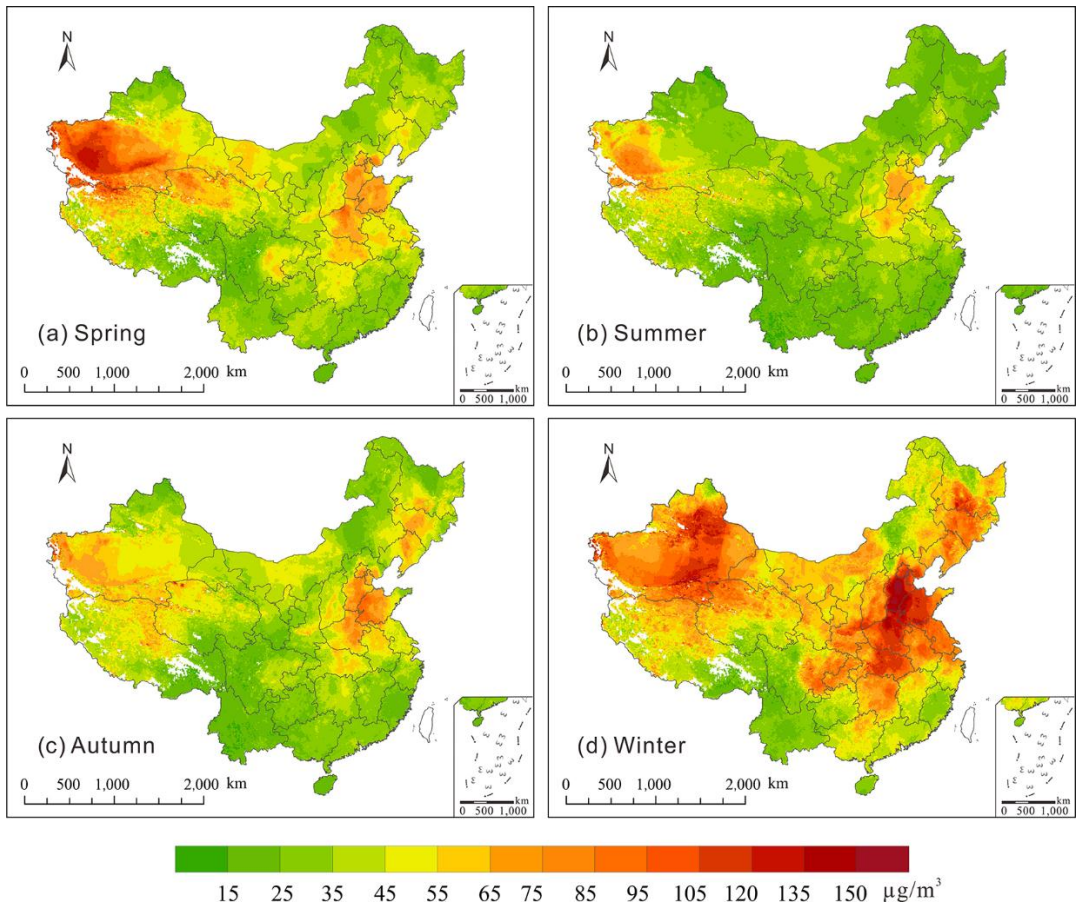

**Figure S3.** Seasonal mean distributions of PM$_{2.5}$ over China in 2015. Spring: March, April, May; Sumer: June, July, August; Autumn: September, October, November; Winter: December, January, February. For prediction at a $0.1° \times 0.1°$ gird, the spatial term is the weighted average of the nearby $n$ grids. The temporal term is obtained from prediction maps only using the input variables and spatial terms ($S\text{-}PM_{2.5}, DIS$). The missing grids of PM$_{2.5}$ predictions are filled by the closest station measurements to calculate the temporal term. White regions indicate missing data.

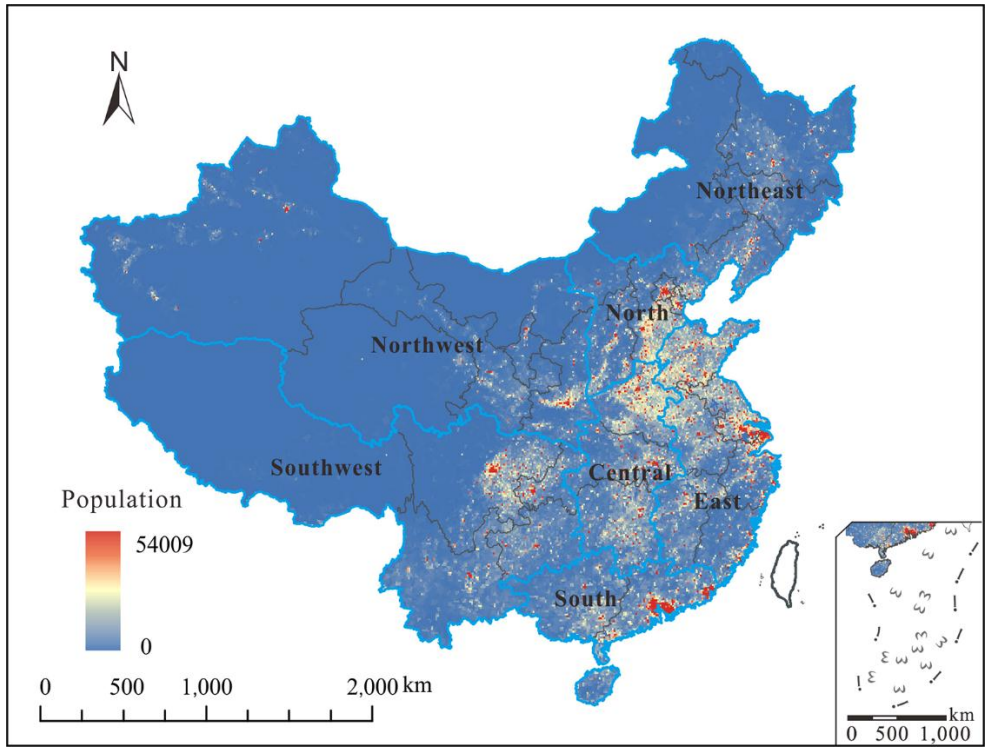

**Figure S4.** Distribution of population in China. Generally speaking, East China has a higher population than West China. In particular, the three megalopolises of Beijing-Tianjin-Hebei (BTH), the Yangtze River Delta (YRD), and the Pearl River Delta (PRD) are the most populous areas in China. The different regions of China are also presented.

| RBM layers | Neurons | Model fitting | | | | Cross-validation | | | |
|---|---|---|---|---|---|---|---|---|---|
| | | R | RMSE | MPE | RPE (%) | R | RMSE | MPE | RPE (%) |
| 1 | 10 | 0.93 | 13.78 | 9.03 | 25.11 | 0.94 | 13.68 | 8.98 | 24.91 |
| | 15 | 0.94 | 13.55 | 8.93 | 24.66 | 0.94 | 13.58 | 8.94 | 24.73 |
| | 20 | 0.94 | 13.66 | 8.99 | 24.90 | 0.94 | 13.52 | 8.92 | 24.61 |
| 2 | 10 | 0.94 | 13.58 | 8.96 | 24.73 | 0.94 | 13.68 | 8.98 | 24.90 |
| | 15 | 0.94 | 13.44 | 8.91 | 24.50 | 0.94 | 13.68 | 9.03 | 24.90 |
| | 20 | 0.94 | 13.17 | 8.82 | 23.96 | 0.94 | 13.69 | 9.03 | 24.93 |
| 3 | 10 | 0.94 | 13.70 | 9.00 | 24.96 | 0.93 | 13.84 | 9.08 | 25.19 |
| | 15 | 0.94 | 13.51 | 8.94 | 24.60 | 0.94 | 13.64 | 8.97 | 24.84 |
| | 20 | 0.94 | 13.44 | 8.91 | 24.47 | 0.94 | 13.68 | 9.04 | 24.91 |

**Table S1.** The performance of the Geoi-DBN model with different parameters.

| Variables | Model fitting | | | | Cross-validation | | | |
|---|---|---|---|---|---|---|---|---|
| | R | RMSE | MPE | RPE (%) | R | RMSE | MPE | RPE (%) |
| a | 0.94 | 13.44 | 8.91 | 24.50 | 0.94 | 13.68 | 9.03 | 24.90 |
| b | 0.94 | 13.08 | 8.77 | 23.80 | 0.94 | 13.57 | 8.96 | 24.71 |
| c | 0.94 | 13.71 | 9.03 | 24.98 | 0.94 | 13.73 | 9.04 | 25.00 |
| d | 0.94 | 13.53 | 8.94 | 24.59 | 0.94 | 13.67 | 8.98 | 24.89 |

a. AOD, meteorological data, NDVI, S-$PM_{2.5}$, T-$PM_{2.5}$, DIS
b. AOD, meteorological data, NDVI, S-$PM_{2.5}$, T-$PM_{2.5}$, DIS, **Road**
c. AOD, meteorological data, NDVI, S-$PM_{2.5}$, T-$PM_{2.5}$, DIS, **Population**
d. AOD, meteorological data, NDVI, S-$PM_{2.5}$, T-$PM_{2.5}$, DIS, **Road, Population**

**Table S2.** Performance of the Geoi-DBN model with the predictors of road and population. Road makes a slight contribution to the improvement of model performance, while population has a negative impact on the Geoi-DBN model.